%% file: main.tex
\documentclass[conference]{IEEEtran}

\usepackage{cite}
\usepackage{amsmath,amssymb,amsfonts}
\usepackage{graphicx}
\usepackage{textcomp}
\usepackage{xcolor}
\usepackage{url}
\usepackage{enumitem}
\usepackage{mathtools}
\usepackage{enumitem}
\usepackage{makecell}
\usepackage{multirow,tabularx}
\usepackage[ruled,linesnumbered]{algorithm2e}
\usepackage{subcaption}
\usepackage[justification=centering]{caption}
\usepackage{hyperref}
\usepackage{flushend}

\usepackage{comment}
\usepackage{ifthen}
\newboolean{showcomments}
\setboolean{showcomments}{true}
\ifthenelse{\boolean{showcomments}}
{ \newcommand{\mynote}[3]{
    \fbox{\bfseries\sffamily\scriptsize#1}
    {\small$\blacktriangleright$\textsf{\emph{\color{#3}{#2}}}$\blacktriangleleft$}}}
{ \newcommand{\mynote}[3]{}}
\newcommand{\shrink}[1]{}
\definecolor{dgreen}{RGB}{46, 124, 49}
\definecolor{purple}{rgb}{0.7,0,0.9}

\newcommand{\eg}{\textit{e.g., }}
\newcommand{\ie}{\textit{i.e., }}

\SetCommentSty{algfont}
\begin{document}

\title{\huge{RISCLESS: A Reinforcement Learning
Strategy to Exploit Unused Cloud Resources}}

\author{\IEEEauthorblockN{SidAhmed Yalles\IEEEauthorrefmark{1}\IEEEauthorrefmark{3}, Mohamed Handaoui\IEEEauthorrefmark{1}\IEEEauthorrefmark{3}, Jean-Emile Dartois\IEEEauthorrefmark{1}\IEEEauthorrefmark{2}, \\ Olivier Barais\IEEEauthorrefmark{1}\IEEEauthorrefmark{2}, Laurent d'Orazio\IEEEauthorrefmark{1}\IEEEauthorrefmark{2}, Jalil Boukhobza\IEEEauthorrefmark{1}\IEEEauthorrefmark{3}}
\vspace{.2cm}
\IEEEauthorblockA{\IEEEauthorrefmark{1}b$<>$com Institute of Research and Technology, 
\IEEEauthorrefmark{2}Univ. Rennes, Inria, CNRS, IRISA,\\ \IEEEauthorrefmark{3}Univ Brest, Lab-STICC, CNRS, UMR 6285, F-29200 Brest, France\\[.2cm]
Email: $\{$sidahmed.yalles,mohamed.handaoui, jean-emile.dartois$\}$@b-com.com,\\ $\{$boukhobza$\}$@univ-brest.fr, $\{$olivier.barais,laurent.dorazio$\}$@irisa.com
}
}

\maketitle

\input{abstract}

\input{introduction}
\input{background}
\input{contribution}
\input{evaluation}
\input{related_work}
\input{conclusion}

\section*{Acknowledgment}
This work was supported by the Institute of Research and Technology b-com, dedicated to digital technologies, funded by the French government through the ANR Investment referenced ANR-A0-AIRT-07.

\bibliographystyle{IEEEtran}
\bibliography{bibliography}
\end{document}

%% file: abstract.tex
\begin{abstract}

One of the main objectives of Cloud Providers (CP) is to guarantee the Service-Level Agreement (SLA) of customers while reducing operating costs. To achieve this goal, CPs have built large-scale datacenters. This leads, however, to underutilized resources and an increase in costs. A way to improve the utilization of resources is to reclaim the unused parts and re-sell them at a lower price. Providing SLA guarantees to customers on reclaimed resources is a challenge due to their high volatility. Some state-of-the-art solutions consider keeping a proportion of resources free to absorb sudden variation in workloads. Others consider stable resources on top of the volatile ones to fill in for the lost resources. However, these strategies either reduce the amount of reclaimable resources or operate on less volatile ones such as Amazon Spot instance. In this paper, we proposed RISCLESS, a Reinforcement Learning strategy to exploit unused Cloud resources. Our approach consists of using a small proportion of stable on-demand resources alongside the ephemeral ones in order to guarantee customers SLA and reduce the overall costs. The approach decides when and how much stable resources to allocate in order to fulfill customers' demands. RISCLESS improved the CPs' profits by an average of 15.9\% compared to state-of-the-art strategies. It also reduced the SLA violation time by an average of 36.7\% while increasing the amount of used ephemeral resources by 19.5\% on average \footnote{This is an extension of the paper published in PDP22 \cite{RISCLESS}}. 

\end{abstract}

\begin{IEEEkeywords}
Cloud, Unused Resources, Ephemeral Resources, Stable Resources, Resource Allocation, SLA, Reinforcement learning, Deep learning
\end{IEEEkeywords}

%% file: introduction.tex
\section{Introduction}
\label{sec:introduction}
Cloud Computing~\cite{cloud_computing_nist}, according to NIST\footnote{\href{https://www.nist.gov/}{NIST: National Institute of Standards and Technology}}, is a model for enabling ubiquitous, convenient, on-demand network access to a shared pool of configurable computing resources. Companies are increasingly moving towards this solution because of its advantages in terms of accessibility, availability, elasticity, flexibility, and reduced costs. However, the expansion of these infrastructures brings to the surface a major problem for CPs, namely the underutilization of resources. In fact, the efficiency of resources, and therefore their profitability, is measured by their degree of utilization: the better the resources are used, the more profitable they are.

In order to cut down the cost of operating underutilized resources, CPs can reclaim the unused part from \textit{regular customers} (the ones who reserved these resources) to (re)sell it at a lower price to other customers (let us call them \textit{ephemeral customers}). These reclaimed resources are by nature volatile. The resale of such resources must meet the ephemeral customers' expectations in terms of SLA. If the SLA is violated, CPs may be subject to penalties. Deploying applications on volatile resources while guaranteeing SLA is still a challenge in the scientific literature~\cite{zhang2016history, dartois2018using, dartois2019cuckoo, javadi2019scavenger, handaoui2020salamander}. Indeed, volatile resources can be lost and returned to their owner (the \textit{regular customers}) in the event that their applications see their resource requirements increase. This change in regular customers' application behavior is very difficult to predict as it may depend on human behavior or interference between co-located applications~\cite{cao2014cpu, fox2009above, dartois2019investigating}.

Different strategies were proposed to improve resource utilization and guarantee customers SLA on ephemeral resources. First, some strategies~\cite{dartois2019cuckoo, handaoui2020salamander,zhang2016history, zhao2020rhythm, patel2020clite, javadi2019scavenger, handaoui2020releaser} solely rely on ephemeral resources. They leave a proportion of those resources unused, called a safety margin, to absorb the sudden increase in regular customers' application demand. The use of a safety margin decreases the amount of reclaimable resources in order to avoid SLA violations. However, when the volatility of resources is significant, these strategies may not perform well. Second, other strategies~\cite{lin2010moon, yan2016tr, yang2017pado, sharma2017portfolio, lin2020backup, ogden2019cloudcoaster} combine stable resources with the volatile ones to guarantee customers' SLA. Nonetheless, they mainly focus on Amazon Spot Instance\footnote{\url{https://aws.amazon.com/ec2/spot/}} which is less volatile than the reclaimed resources. The customers also receive a notification from Amazon prior to the actual interruption. In the case of ephemeral resources, such a notification does not exist. Thus guaranteeing SLA while increasing the CPs' profits is a real challenge.

We argue in this paper that Machine Learning (ML) algorithms can be used to determine when and how many stable resources to allocate on top of the ephemeral ones. Specifically, we used Reinforcement Learning (RL) algorithm, a field of ML used for decision making. The main reason for using RL is due to the limitations of classical solutions. In fact, most of the solutions~\cite{dartois2019cuckoo, lin2010moon, yan2016tr} are centered on the parametric improvement and optimization of allocation strategies based on heuristics. The algorithms are sometimes difficult and time-consuming to (re)configure. Above all, the solutions are rigid and not flexible to changes with regards to the environment (\eg workloads' requirements). Thus, it compromises their performance and usability on the Cloud. To solve this problem, several researchers have proposed methods mainly based on ML with models that are capable of autonomously learn resource allocation policies. More specifically, a number of RL approaches have been proposed for task scheduling~\cite{chen2017deep, li2019deepjs, liang2019job, ye2018new} and resource allocation~\cite{dutreilh2011using, galstyan2004resource, liu2017hierarchical}. Although these studies do not consider ephemeral resources, they show that RL is indeed a promising choice to solve similar problems.

This paper proposes a new approach to Cloud resource allocation that improves the utilization of ephemeral resources while guaranteeing SLA. The proposed solution computes the volatility rate of resources using past utilization traces. It then captures a set of information, namely \textit{i)} customers allocation request, \textit{ii)} the amount of stable and ephemeral resources available and \textit{iii)} the volatility rate of resources. The information is used for the decision process of when to allocate ephemeral and stable resources in order to respond to customer's requests. All while increasing CPs' profits because stable resources are more expensive than the ephemeral ones. The solution also aims to reduce the possible penalties for violating the SLA.



The experimental evaluation was based on Amazon's Spot Instance\footnote{https://aws.amazon.com/ec2/spot/pricing/} model prices. The results show that our solution allows reducing SLA violation time by 36.7\% on average compared to the safety margin strategies. The use of stable resources allows RISCLESS to compensate for the possible loss of allocated ephemeral resources. The solution provides, in most cases, the amount of resources requested by customers. The reduction in violation time also proves that the model estimates more precisely the amount of volatile resources lost at each time step.

The remainder of this paper is organized as follows. Section~\ref{section:background} provides a background about RL and Deep ML. Section~\ref{section:contribution} details our contribution. Then, Section~\ref{section:evaluation} describes the experimental evaluation and the results obtained. In Section \ref{section:related_work}, we discuss the related work. Finally, Section~\ref{section:conclusion} concludes the paper and discusses future work.

%% file: background.tex
\section{background}
\label{section:background}
In this section, we briefly describe the RL technique. We also define Deep Machine Learning that is used by several RL algorithms for solving large and/or complex problems.

\subsection{Reinforcement Learning}
\begin{figure}
\centering
\includegraphics[width=0.35\textwidth]{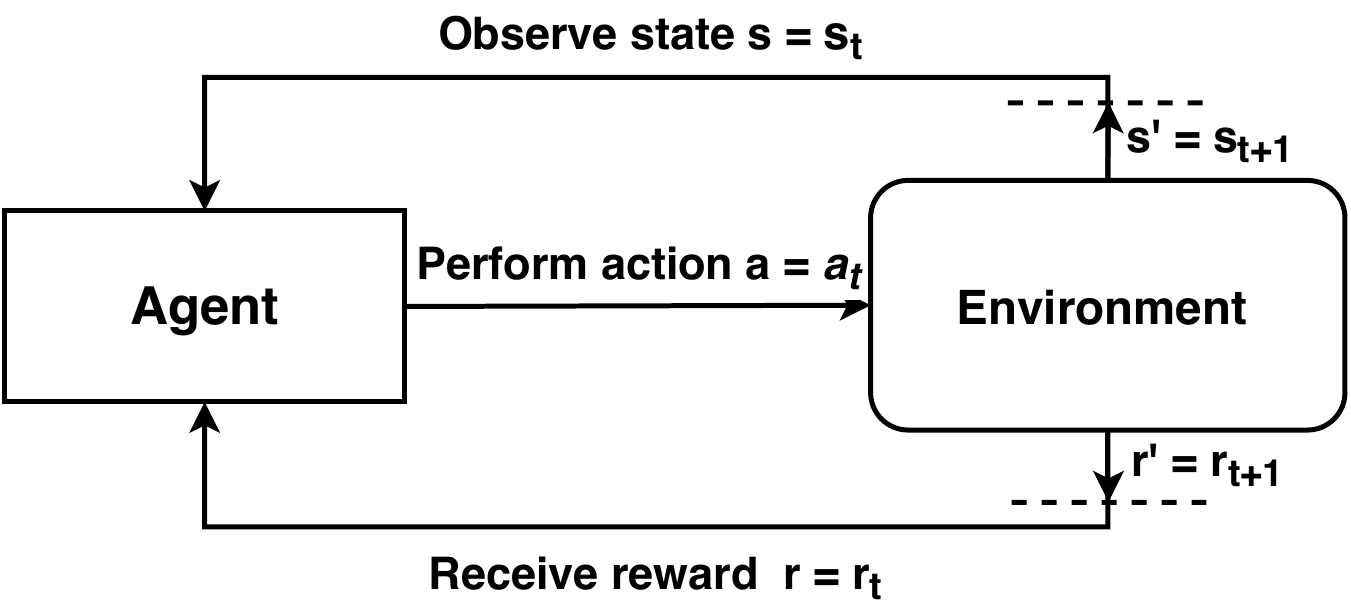}
\caption{Reinforcement Learning Architecture Overview}
\label{fig:reinfrocement_learning}
\end{figure}

RL~\cite{geron2019hands} is a field of ML alongside supervised learning and unsupervised learning. In RL, an agent (model) is responsible for learning a policy (\ie behavior) through trial and error. The agent tries to find the best policy that allows choosing the right decisions through experience. In order to gather experience, the agent interacts with its environment. 

Fig.\ref{fig:reinfrocement_learning} illustrates the basic principle of RL. At each time step, the agent observes the state of the environment $s_t \in S$ and chooses one of the possible actions $a_t \in A(s_t)$. By executing the action of the agent, the environment goes from the state $s_t$ to the state $s_{t+1}$. The agent receives, in addition to the new state $s_{t+1}$, a reward $r_{t+1}$ that is a feedback on the quality of its action $a_t$ to the state $s_t$.

RL is formalized through a Markov Decision Process (MDP)~\cite{bellman1957markovian}. The Markov hypothesis assumes independence between past and future states, which means that the state of the environment at instant $t+1$ depends only on its state and the action at time $t$. An MDP is represented by a quadruplet $<S, A, T, R>$:

\begin{itemize}[leftmargin=5ex]
    \item $S$: set of the environment states.
    \item $A$: set of actions.
    \item $T_{s,a}^{s'}:S \times A \times S \to [0,1]$ : the probability to have a transition from state $s$ to $s'$ by performing action $a$ at time $t$.
    \item $R_{s,a}^{s'}:S \times A \times S \to \mathbb{R}$ : defines the reward function of the agent when transitioning from state $s$ to $s'$ by performing action $a$ at time $t$.
\end{itemize}

A policy noted $\pi:S \to A$, maps actions to states. In each state, the agent selects, according to the policy, the action that maximizes the future expected rewards. Optimizing the policy involves optimizing one of the two functions:
\begin{itemize}
    \item Value function $V_\pi(s)$: it represents the expected future rewards if the agent follows the policy $\pi$ from state $s$.
    \item Q-value function $Q_\pi(s,a)$: it represents the expected future rewards following $\pi$ from state $s$ and performing action $a$.
\end{itemize}




Several algorithms~\cite{geron2019hands} exist for optimizing a policy such as Dynamic Programming, Monte Carlo, and TD learning. However, when the state space (or action space) is too large, approximation techniques are used such as Deep Machine Learning~\cite{goodfellow2016deep}.

\subsection{Deep Machine Learning}

Deep ML~\cite{goodfellow2016deep} is defined by what is called artificial neural networks. Each artificial neuron has $n$ input connections and a weight $\theta$.
Neural networks estimate a non-linear function $f^*(x)$ by combining several neurons. The set of parameters $\theta$ that contains all the weights $\theta_{i}$ of all neurons must be adjusted so that the neural network gives the best possible approximation of the function. Each layer has an activation function that defines the output of each neuron. The process of finding a good set of $\theta$ parameters is referred to as model learning.

A neural network organizes artificial neurons into different layers. The neurons of these layers are connected to each other in a retro-propagated way. There are no connections that are sent back to previous neurons. Each network has an input layer, which processes the raw input data and an output layer, which contains the approximate result. Between these two layers, there may be one or more so-called hidden layers where calculations are performed.

%% file: contribution.tex
\section{RISCLESS : A ReInforcement Learning Strategy to Guarantee SLA}
\label{section:contribution}

In this section, we describe the proposed Reinforcement Learning strategy that allocates resources according to customers' requests. The solution combines volatile and stable on-demand resources to reduce the cost of allocation. At the same time, the solution should minimize SLA violations.

\subsection{Architecture overview}
\begin{figure}
    \centering
    \includegraphics[width=0.45\textwidth]{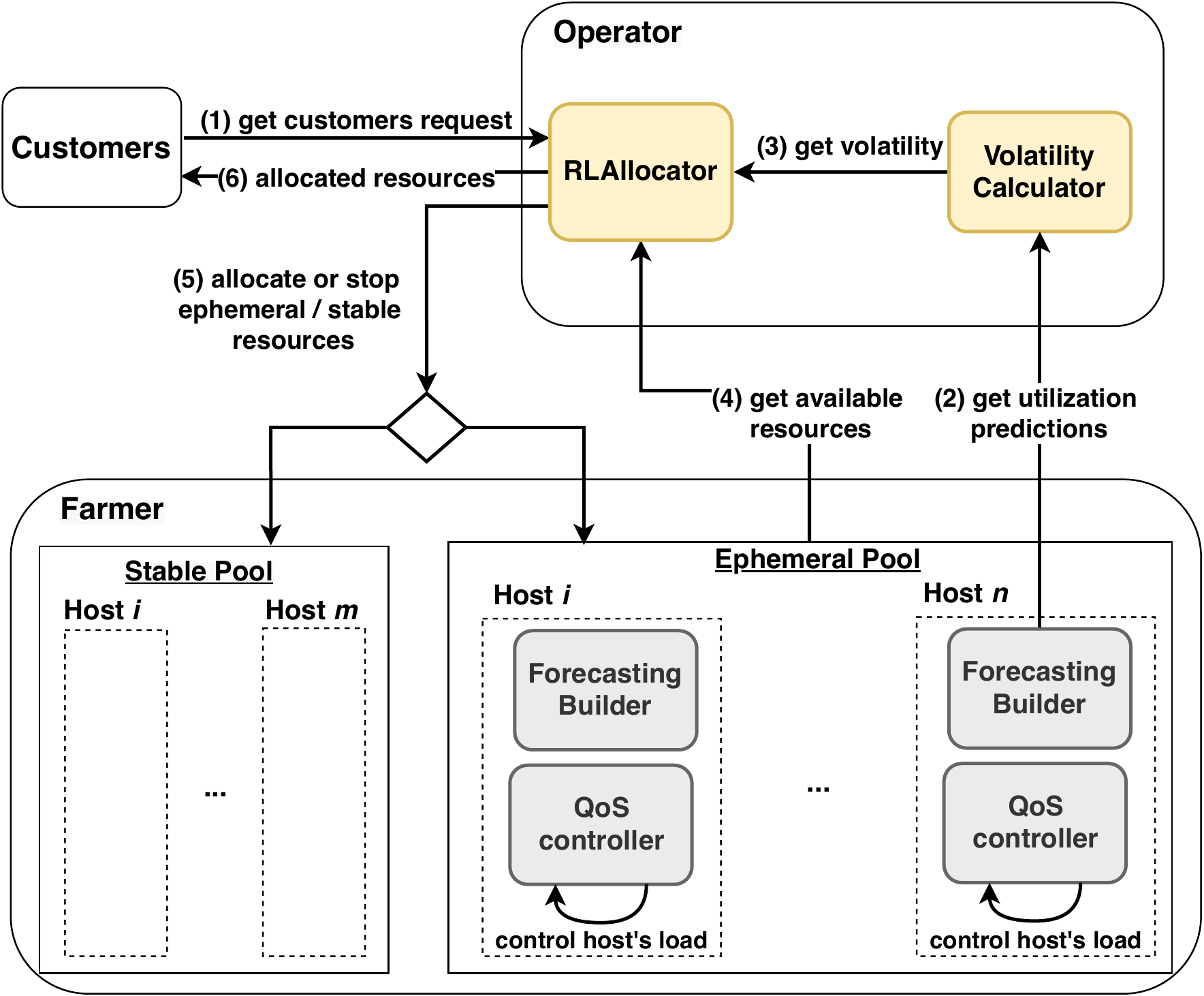}
    \caption{Overview architecture that deploys RISCLESS modules }\label{fig:architecture} 
\end{figure}

Fig.\ref{fig:architecture} presents an overview of the architecture that deploys our solution called RISCLESS (\textbf{R}e\textbf{I}nforcement Learning \textbf{S}trategy to Guarantee SLA on \textbf{CL}oud \textbf{E}phemeral and \textbf{S}table Re\textbf{S}ources). There are three main actors:
\begin{enumerate}[leftmargin=4ex]
    \item \textbf{Farmers}: datacenter owners, that seek to reduce their Total Cost of Ownership (TCO) by offering unused resources to customers. We suppose that these farmers have stable resources that could be allocated on-demand with higher costs compared to the unused resources. 
    \item \textbf{Customers}: we focus here on customers that request ephemeral cloud resources at a lower cost (\ie ephemeral customers).
    \item \textbf{Operator}: the interface between farmers and customers. They aim at minimizing farmers' TCO by offering unused resources to customers with SLA requirements.
\end{enumerate}

Globally, the architecture is composed of two pools of resources namely ephemeral and stable resources. Each of the two resource pools is formed by a set of hosts. The hosts forming the ephemeral pool deploy the following two modules:
\begin{itemize}
    \item \textbf{Forecasting Builder}: designed by the authors of ~\cite{dartois2018using}, this module is responsible for predicting the next 24 hours of future resource utilization on a host-level. The amount of available resources is then computed according to host capacities. 
    \item \textbf{QoS Controller}: designed by the authors of ~\cite{dartois2019cuckoo, handaoui2020salamander}, it ensures that the SLA of regular customers who have reserved resources on the host are respected. It does so by reducing the utilization of ephemeral resources if their owners (\ie regular customers) need them.
\end{itemize}

Finally, the Operator needs to deploy two modules representing our contribution: 
\begin{itemize}
    \item \textbf{Volatility Calculator}: this module is used to reduce the complexity of the resource allocation process. It summarizes the multiple points (\ie 24 hours) predictions from the Forecasting Builder into a single value. This value represents the volatility rate (\ie probability) of losing the ephemeral resources. 
    
    
    \item \textbf{RLAllocator}: represents the decision-maker of our solution. It is based on an RL algorithm to decide when to allocate ephemeral and stable resources. The module aims at allocating the resources requested by customers with SLA requirements while maximizing CPs' profits. 
\end{itemize}

The architecture in Fig.\ref{fig:architecture} illustrates the workflow and communication between the different actors and modules. The workflow starts with (1) the customers requesting resources for allocation through the \textit{Customers} interface. The request is received by the \textit{Operator} who transfers it to the \textit{RLAllocator} module. The \textit{Volatility Calculator} module then (2) retrieves the predictions of ephemeral resources and calculates their volatility rate. The module sends (3) the volatility rate to the \textit{RLAllocator} module. The \textit{RLAllocator} module also (4) receives the available ephemeral resource capacities. Once all the data have been collected, the \textit{RLAllocator} module (5) decides on the pool of resources to allocate namely ephemeral or stable resources. Finally, The Operator (6) informs the customers of the allocated resources.

\vspace{0.3cm}
We detail in what follows the design of the two modules presented above that is the Volatility Calculator and RLAllocator.

\subsection{Volatility Calculator}
As mentioned above, the Forecasting Builder predicts future resource utilization for every host. Having the hosts' predictions for the next 24 hours as input for the RLAllocator is costly as the size increases with the number of hosts. Through the Volatility Calculator module, our goal is to reduce the verbose information and only provide a single value that represents the volatility rate of losing resources in the ephemeral pool.

The volatility rate provides a summary of prediction errors to the RLAllocator module. The latter can then make allocation decisions based on the provided information. The module receives as input both the past prediction and utilization of resources during a $\Delta$t window (\eg 24 hours). The module then computes the volatility rate $p\in[0,1]$ of the ephemeral pool. Finally, the module outputs the value for the RLAllocator module.


\begin{table}
    \renewcommand{\arraystretch}{1.2}
    \setlength{\tabcolsep}{2.5pt}
    
    \caption{Example of the utilization and prediction traces from the Forcasting Builder module}
    \label{tab:utilization_example}
    \begin{tabular}{c|c|c|c|c|c|c|c}
        \hline
        \textbf{$t$} & \textbf{$\widehat{y}_{cpu}$} & \textbf{$y_{cpu}$} & \textbf{$e_{cpu}=\widehat{y}_{cpu}-y_{cpu}$} & \textbf{$\widehat{y}_{m}$} & \textbf{$y_{m}$} & \textbf{$e_{m}=\widehat{y}_{m}-y_{m}$} & \textbf{$z$}\\ 
        \hline
        0 & 30\% & 60\% & -30\% & 40\% & 60\% & -20\% & 1\\ 
        1 & 40\% & 30\% & 10\% & 53\% & 50\% & 3\% & 0\\ 
        \vdots & \vdots & \vdots & \vdots & \vdots & \vdots & \vdots & \vdots \\
        479 & 41\% & 52\% & -11\% & 45\% & 40\% & 5\% & 1\\ 
        \hline
    \end{tabular}
\end{table}

Table \ref{tab:utilization_example} shows an example of traces with predictions from the Forecasting Builder for a time window of $\Delta$t=24h with a 3-minute sampling period. It contains the following for both the CPU and memory metrics:
\begin{itemize}
    \item Actual measures of utilization: $y_{metric}$
    \item Predictions of future resource utilization: $\widehat{y}_{metric}$
    \item Prediction errors: $e_{metric}=\widehat{y}_{metric}-y_{metric}$
\end{itemize}


The volatility rate represents the probability of underestimating the amount of resource utilization. In other words, resources are lost if $\widehat{y}_{metric} < y_{metric}$ and the amount of resources lost is proportional to the prediction error.

To calculate this probability, a random variable $z$ is used where $z_{t}$ represents whether the predictions underestimated the CPU or memory usage at time $t$. 
Table \ref{tab:utilization_example} shows the values that the variable $z$ takes according to its definition. It is set to '$1$' if the prediction is underestimated (\ie $e_{metric} < 0$). For example, at $t=0$, the predicted CPU is $\widehat{y}=30\%$ but the measured utilization is $y=60\%$, thus the prediction error is $e_{cpu} = -30\% < 0$ and $z=1$. Note that for this first version, the size of the prediction error does not impact the value of $z$ (\eg errors of -$1\%$ or -$50\%$  $\implies z=1$).

Assuming that the different measures are independent of each other, the variable $z$ follows a Bernoulli's distribution\footnote{The Bernoulli's distribution, or Bernoulli's law~\cite{stigler1986history}, is a discrete probability distribution, which takes the value 1 with the probability $p$ and 0 with the probability $q = 1 - p$.} of parameter $p$. With $p$ representing the probability of underestimation:
\begin{align}
    & z_{t} \rightsquigarrow B(p)\\
    & P_\text{\footnotesize{underestimation}} = P_(\Delta t) = p
\end{align}

\noindent To estimate $p$, an empirical estimator $\widehat{p}$ is used. The estimator is the mean over a $\Delta t$ time window of the $z$ values as follows: 
\begin{IEEEeqnarray}{lCr}
	\widehat{p} = \frac{\sum_{t \in \Delta t} z_{t}}{\Delta t}
\end{IEEEeqnarray}

In summary, the volatility rate is the probability of losing resources. It is computed on the basis of the underestimation part of prediction errors. Note that the resource volatility is recalculated at the beginning of each $\Delta$t period by the Volatility Calculator module. This value is later provided to the RLAllocator as an indicator of resource volatility. 


\subsection{RLAllocator}
The RLAllocator module is the decision-maker in the solution architecture. Its objective consists of deciding when to allocate ephemeral and stable resources in order to maximize CPs' profits and reduce SLA violations.


This module was built using RL. As explained in Section~\ref{section:background}, it is formalized with the MDP composed of the quadruplet $<S,A,T,R>$ with: $S$ being the set of states of the environment, $A$ the set of actions, $T$ the transition function, and $R$ the reward function.

\subsubsection{\textbf{Environment}}
Each of the two pools of resources is characterized by its available capacity in terms of CPU and memory (other resources could be considered). The ephemeral pool is further characterized by their volatility rate. Both resource types have a cost of allocation. Ephemeral resources are less expensive than the stable ones with up to 90\% difference in the case of Amazon Spot Instance~\footnote{\url{https://aws.amazon.com/ec2/spot/}}. 


\subsubsection{\textbf{State space $\mathbf{S}$}} \label{sec:state_space}
At each time step $t$, the state is characterized by the customers' request, the quantity of ephemeral and stable resources allocated, the available capacity of the two types of resources, and finally the volatility rate of the ephemeral resources. The state of the environment is defined as follows:


\begin{multline*}
    S =  \{res_{rem},\; res_{alloc}(e),\; res_{alloc}(s),\; res_{avail}(e),\; \\ res_{avail}(s),\; p\}
\end{multline*}

With:
\begin{itemize}
        \item $ res_{rem} \in \mathbb{N} $: the amount of the remaining resources to be allocated. 
        \item $ res_{alloc}(e) 	\in \mathbb{N} $: the amount of ephemeral resources allocated.
        \item $  res_{alloc}(s)	\in \mathbb{N} $: the amount of stable resources allocated.
        \item $  res_{avail}(e)  \in \mathbb{N} $: the amount of available ephemeral resources.
        \item $  res_{avail}(s)  \in \mathbb{N} $: the amount of available stable resources.
        \item $ p \in [0,1] $: the volatility rate.
\end{itemize}

Note that the model can allocate more resources than originally requested. This is permissible for fast recovery when losing resources to avoid SLA violations.


\subsubsection{\textbf{Action space $\mathbf{A}$}} \label{sec:action_space} At each time step $t$, the model can perform an action $a$. The set of actions we defined are:

$A = \{ a_1, a_2, a_3, a_4, a_5 \}$

with:
\begin{itemize}
    \item $a_1$: Allocate an ephemeral resource unit.
    \item $a_2$: Remove an ephemeral resource unit.
    \item $a_3$: Allocate a stable resource unit.
    \item $a_4$: Remove a stable resource unit.
    \item $a_5$: Do nothing.
\end{itemize}

A resource unit is defined as an amount of vCPU and memory that are allocated at the same time to a customer (\eg a resource unit of 2 vCPU and 8 Gb). 

\subsubsection{\textbf{Reward function $\mathbf{R}$}} Our goal is to maximize CPs' profits from selling ephemeral resources and minimize the cost of stable resources and SLA violations. The reward function is defined for each state as: 
\begin{multline}
    r = res_{alloc}(e) \times CPE - res_{alloc}(s) \; \times \; \\ CPS - res_{rem} \times CPV
\end{multline}
with :
\begin{itemize}
    \item $CPE$: cost per ephemeral resource unit.
    \item $CPS$: cost per stable resource unit.
    \item $CPV$: cost of SLA violation penalty.
\end{itemize}

Each ephemeral or stable resource has a cost per unit $CPE$ and $CPS$. While the SLA violation has a cost per violation $CPV$. In a state S, the reward function considers the total cost of the amount of ephemeral resources allocated $res_{alloc}(e)$ which is considered as the profit. The amount of stable resources $res_{alloc}(s)$ has to be minimized since its cost is higher than the ephemeral ones. SLA violation can occur in two cases: i) when losing an ephemeral resource, ii) when the customer requests are not provided. In both cases, the remaining resource to allocate $res_{rem}$ increases, hence increasing SLA violation penalties.


\subsubsection{\textbf{Model algorithm}} In order to solve the Cloud resource allocation problem, we train the RL agent using the Deep Q-Network (DQN) algorithm~\cite{mnih2015human}. DQN is used to approximate the Q-values (defined in Section \ref{section:background}) using neural networks with a single function (called Q network). Since the state representation of the allocation problem is too large (see Section \ref{sec:state_space}), DQN can approximate values for the Cloud states that have never been encountered during the learning process. Algorithm \ref{algo:dqn} represents the pseudo-code of allocating Cloud ephemeral and stable resources using DQN.


The algorithm starts by initializing (line 1) the configuration of the agent's model (\eg the architecture of the neural network, see Section \ref{sec:implementation} for the initialization used in the evaluation). Then, it initializes a buffer that stores previous resource allocation experiences (line 2). The buffer is used to improve the learning process of the agent. Each experience contains the state (\eg allocated amount of ephemeral and stable resources), action (\eg allocate stable resource), reward (\eg SLA violation penalty), and the next state of the Cloud environment. 
The agent starts by receiving the amount of resources to allocate (line 3). Then it receives (lines 5-7) the available amount of resources for the ephemeral and stable pools. The agent then makes either a random resource allocation decision (\ie action) or the best one according to a probability $\epsilon$ (line 8). The random selection of actions is necessary since initially, the agent does not have any previous experience. Note that the probability $\epsilon$ is reduced during the evaluation in order to assess the learned strategy. Then, the selected action is sent to the Cloud environment. Afterward, the agent fetches both the reward and the new state of the environment (line 9). This current resource allocation experience is stored in the buffer (line 10) which is used for the learning process (lines 11-13). The decision process is repeated while resources remain to be allocated for customers.

\SetNlSty{textbf}{}{.}
\begin{algorithm}
    \SetAlgoLined
    \SetNoFillComment
    \small

    agent = initialize\_DQN\_model()\;
    experiences = initialize\_experience\_buffer()\;
    
    $res_{rem}$ = get\_remaining\_resources\_to\_allocate()\;
    \While(\tcp*[h]{remaining resources to allocate}){$res_{rem} >$ 0}{
        $res_{alloc}$ = get\_allocated\_resources()\;
        $res_{avail}$ = get\_available\_resources()\;
        p = compute\_volatility\_rate()\;
        
        select $a$= $\left\{
            \begin{array}{ll}
                \mbox{random action} & \mbox{with probability $\epsilon$}\\
                \mbox{best action} & \mbox{else}
            \end{array}
        \right.$
        
        
        reward = observe\_reward\_value()\;
        
        experiences.add\_current\_experience()
        
        \If{should\_update}{
            agent.update(experiences)\;
        }
        
        $res_{rem}$ = get\_remaining\_resources\_to\_allocate()\;
    }
  \caption{Pseudo-code of allocating Cloud ephemeral and stable resources using DQN}
  \label{algo:dqn}
\end{algorithm}

%% file: evaluation.tex
\section{Experimental evaluation}
\label{section:evaluation}

In this section, we present the results of the experimental evaluation of RISCLESS. Through the experiments, we try to answer the following Research Questions (RQ):

\begin{itemize}
    \item \textbf{RQ1:} What is the overall performance of RISCLESS in terms of resource utilization, SLA violations, and CPs' profits?
    \item \textbf{RQ2:} How many on-demand stable resources does RISCLESS use on top of the ephemeral ones to reduce SLA violations and to increase profits ?
\end{itemize}

We evaluate RISCLESS by comparing it to other resource allocation approaches that exploit ephemeral resources. The comparison is done using in-production traces of three datacenters (presented below). The results are then analyzed in terms of stable resources percentage compared to the total allocated by RISCLESS.

\subsection{Experimental setup}
\subsubsection{\textbf{Datasets}} 
\begin{figure}[htbp]
    \centering
    \begin{subfigure}[b]{0.24\textwidth}
      \centering
      \includegraphics[width=1\textwidth]{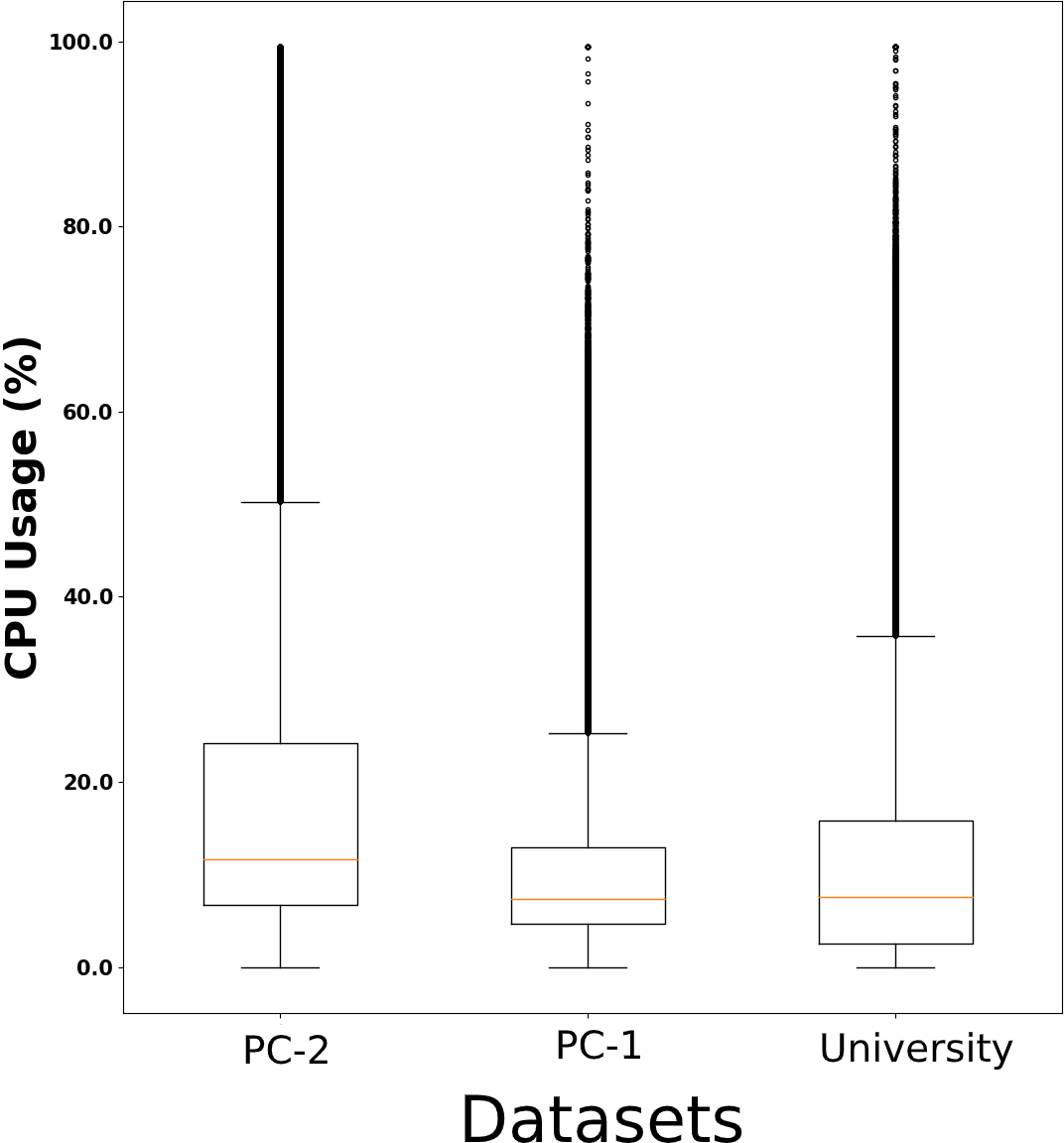}
      \caption{CPU utilization}
      \label{fig:average_utilization_cpu}
    \end{subfigure}
    \begin{subfigure}[b]{0.24\textwidth}
      \centering
      \includegraphics[width=1\textwidth]{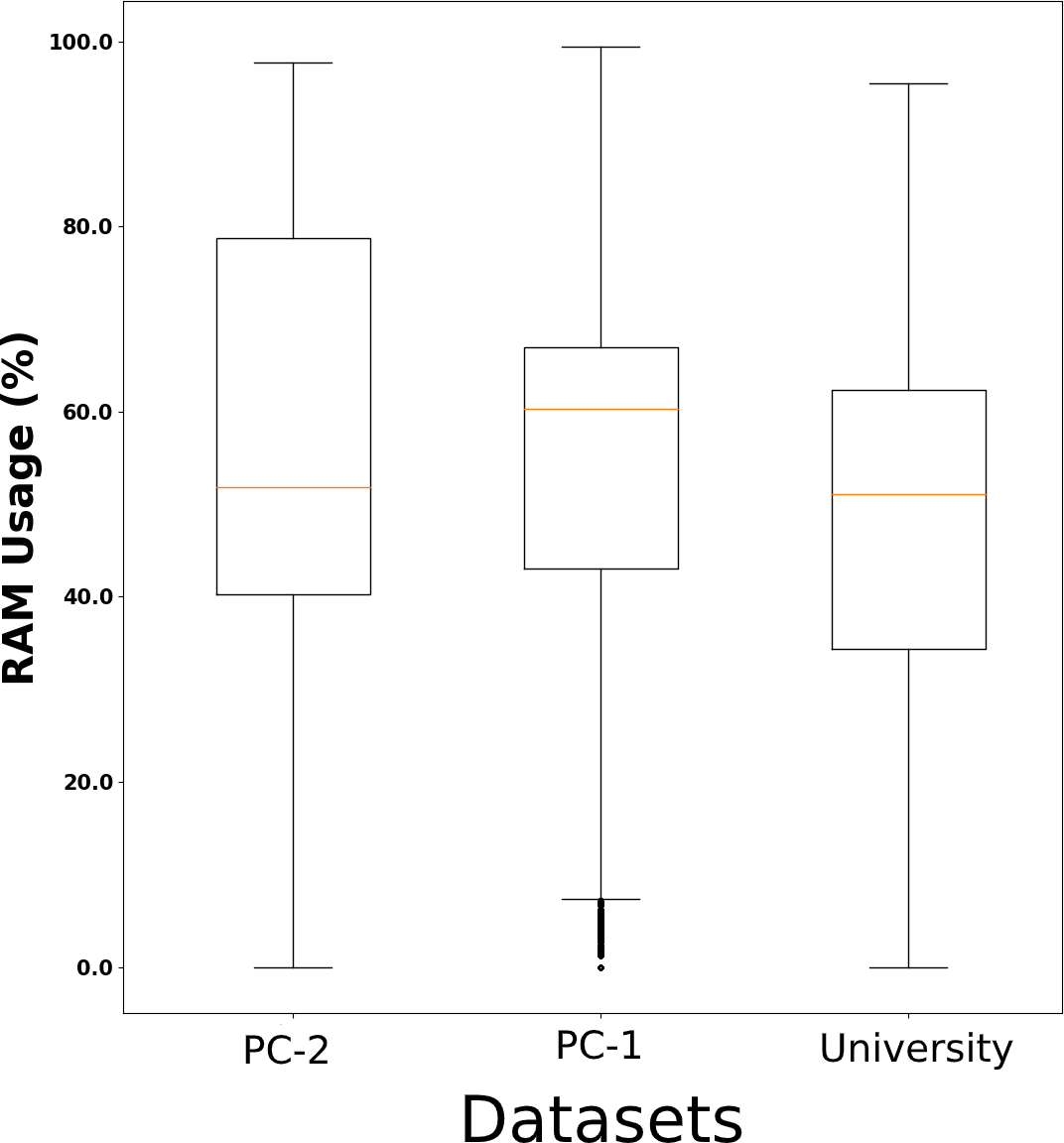}
      \caption{RAM utilization}
      \label{fig:average_utilization_ram}
    \end{subfigure}
    
    \caption{Average resources utilization for the datasets} 
    \label{fig:average_utilization}
\end{figure}

the traces used for training as well as for evaluation are from three different datacenters. One datacenter is from a University, and two are from Private Companies labeled PC-1 and PC-2 respectively. The total capacity of each datacenter in terms of hosts, total CPU power, and the amount of memory is represented in Table \ref{tab:datacenters_capacity}. Fig.\ref{fig:average_utilization} is a boxplot that shows the resource utilization for the CPU and RAM for each datacenter. The traces used are recorded over a period of 6 consecutive months with a 3-minute sampling period.

\begin{table}
  \renewcommand{\arraystretch}{1.3}
  \setlength{\tabcolsep}{15pt}
  \begin{center}
    \caption{Summary information of each datacenter}
    \label{tab:datacenters_capacity}
    \begin{tabular}{cccccc}
      \hline
      \thead{Datacenter} & \thead{Number\\of hosts} & \thead{CPU \\ (cores)}  & \thead{RAM \\(TB)} \\ 
      \hline
      PC-1 & 9 & 120 & 1.2\\
      PC- 2 & 27 & 230 & 3.8 \\
      University & 6 & 72 & 1.5\\
      \hline
    \end{tabular}
  \end{center}
\end{table}

\subsubsection{\textbf{Customers requests}}
\label{sec:requests_assumption}
At each time step, we suppose that the requested resources by customers can utilize all the available unused resources. This means that if 14.6\% of CPU is used then the customers' request is 85.4\% CPU provided that RAM is available. We do this in order to evaluate the maximum of reclaimable resources in a datacenter with the least of SLA violations, which should result in higher profits for the CPs.

\subsubsection{\textbf{Resource allocation approaches}}
RISCLESS is compared to a set of resource allocation approaches operating in the same context (\ie ephemeral resources with SLA guarantees). These approaches use the safety margin method to reduce SLA violations. Other approaches that combine stable and ephemeral resources are not comparable to ours since they are tied to a specific type of application such as data processing (more details in Section \ref{section:related_work}). The following provides a brief description of the different approaches:
\begin{itemize}[leftmargin=4ex]
    \item \textbf{Fixed:} proposed by the authors of~\cite{dartois2019cuckoo}. This approach is a static safety margin percentage of 5\% selected empirically from different datasets.
    
    \item \textbf{Scavenger}: proposed by the authors of~\cite{javadi2019scavenger}. This approach uses the mean and standard deviation of resource utilization history to compute a dynamic safety margin.
    
    \item \textbf{ReLeaSER}: proposed by the authors of~\cite{handaoui2020releaser}. This approach is based on Reinforcement Learning that selects a dynamic safety margin according to the resource prediction errors.
\end{itemize}

For these strategies that use safety margins, we suppose that customers' requests could be satisfied only by ephemeral resources.

\subsubsection{\textbf{Implementation}}\label{sec:implementation}

The experiments were performed offline on an ad-hoc simulator with in-production traces. RISCLESS was implemented using Keras~\cite{chollet2015keras} v.~$2.3.1$, a framework that facilitates the development of deep machine learning models. Keras is an abstraction of Google's open-source framework TensorFlow~\cite{abadi2016tensorflow}. We used TensorFlow GPU v.$~1.14.0$. The architecture of the agent's neural network is as follow:
\begin{itemize}[leftmargin=4ex]
    \item Neural network architecture~\cite{goodfellow2016deep}:
    \begin{itemize}[leftmargin=2ex] 
        \item 2 dense layers, 24 neurons (state size), ReLu activation,
        \item 1 dense layer, 5 neurons (actions size), Linear activation.
    \end{itemize}
    \item Mean Square Error as the error function: \\$mse = \frac{1}{n}\sum_{i=1}^{n}(y_i-\hat{y_i})^2$,
    \item Batch size (number of transitions): $batch\_size = 50$,
    \item Learning rate: $\alpha = 0.001$,
    \item Discount factor: $\gamma = 0.95$,
    \item Agent exploration decay: $\epsilon = \epsilon * 0.995$,
    \item Replay memory (number of transitions): $memory = 20000$,
\end{itemize}

\noindent The model training takes place over several days. The training is performed on 80\% of PC-2 traces, while 20\% are used for testing. Note that the environment is reset at the beginning of each episode.

\subsection{Evaluation metrics}
For evaluating the performance of RISCLESS and compare it with the different approaches, we used the following metrics:

\begin{enumerate}[leftmargin=4ex]
    \item \textbf{Total profits:} we measure the total CPs' profits by the cost related to the profit of ephemeral resources sold (for all the hosts of the datacenter). We take into account the cost of SLA violations and the cost related to the on-demand stable resources we needed to use for SLA sake. The cost of stable resources is not considered for the approaches that solely rely on a safety margin. Thus, the profits are computed as follows:
    \vspace{-0.2cm}
    \begin{multline}
	    profits = ephemeral \; resources \; profit \; -\\[-.1cm]
	              stable \; resources \; cost - SLA \; cost
    \end{multline}
    
    \item \textbf{SLA violation time:} it indicates the cumulative time during which the SLA was violated. The SLA is violated when the Operator does not provide the resources that the customers have requested. This may occur mainly when ephemeral resources are lost.
    
    \item \textbf{Amount of reclaimed ephemeral resources:} it measures the cumulative amount of ephemeral resources that were used without affecting the SLA of customers. 
\end{enumerate}

The evaluation of the first two metrics is based on a real economical model from Amazon AWS. The economical model comprises of resources cost according to their type (\ie volatile or stable) and the penalty model for violating SLA:

\begin{enumerate}[leftmargin=4ex]
    \item \textbf{Resource costs:} we based our resource costs on Amazon AWS instance type \textit{t2.large}\footnote{\url{https://aws.amazon.com/ec2/instance-types/t2/}} that corresponds to 2 vCPU and 8 Gb of memory:
    \begin{itemize}[leftmargin=4ex]
        \item Cost of an ephemeral instance: 0.0317 \$/hour
        \item Cost of a stable instance: 0.0928 \$/hour
    \end{itemize}
    \item \textbf{SLA violation:} the penalty model used calculates the cost of violating the SLA as a discount on the profit related to the sold instances. The discount percentage is based on the cumulative violation time over one day as follows:
    \begin{itemize}[leftmargin=4ex]
        \item Between 15 and 120 minutes: 10\% discount
        \item Between 120 and 720 minutes: 15\% discount
        \item More than 720 minutes: 30\% discount
    \end{itemize}
\end{enumerate}

\subsubsection{\textbf{RQ-1. Overall performance of RISCLESS}} This test focuses on the performance of RISCLESS compared to the other approaches. The performance is evaluated in relation to i) the total CPs' profits, ii) the cumulative SLA violation time, and iii) the amount of reclaimed ephemeral resources.

\begin{figure*}[htbp]
    \centering
    \begin{subfigure}[b]{0.32\textwidth}
      \centering
      \includegraphics[width=0.9\textwidth]{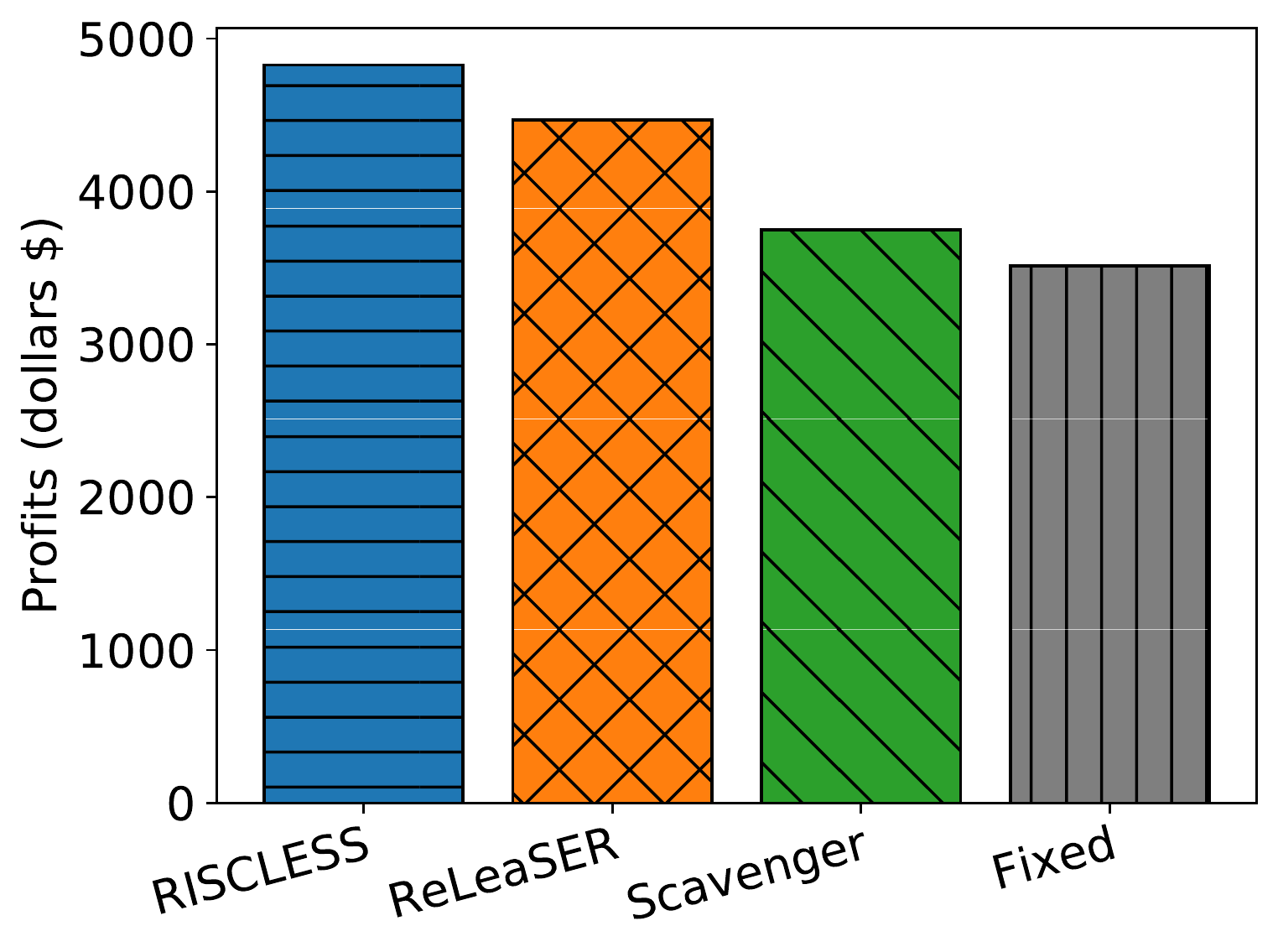}
      \caption{Private Company 1}
      \label{fig:savings_pc1}
    \end{subfigure}
    \begin{subfigure}[b]{0.32\textwidth}
      \centering
      \includegraphics[width=0.9\textwidth]{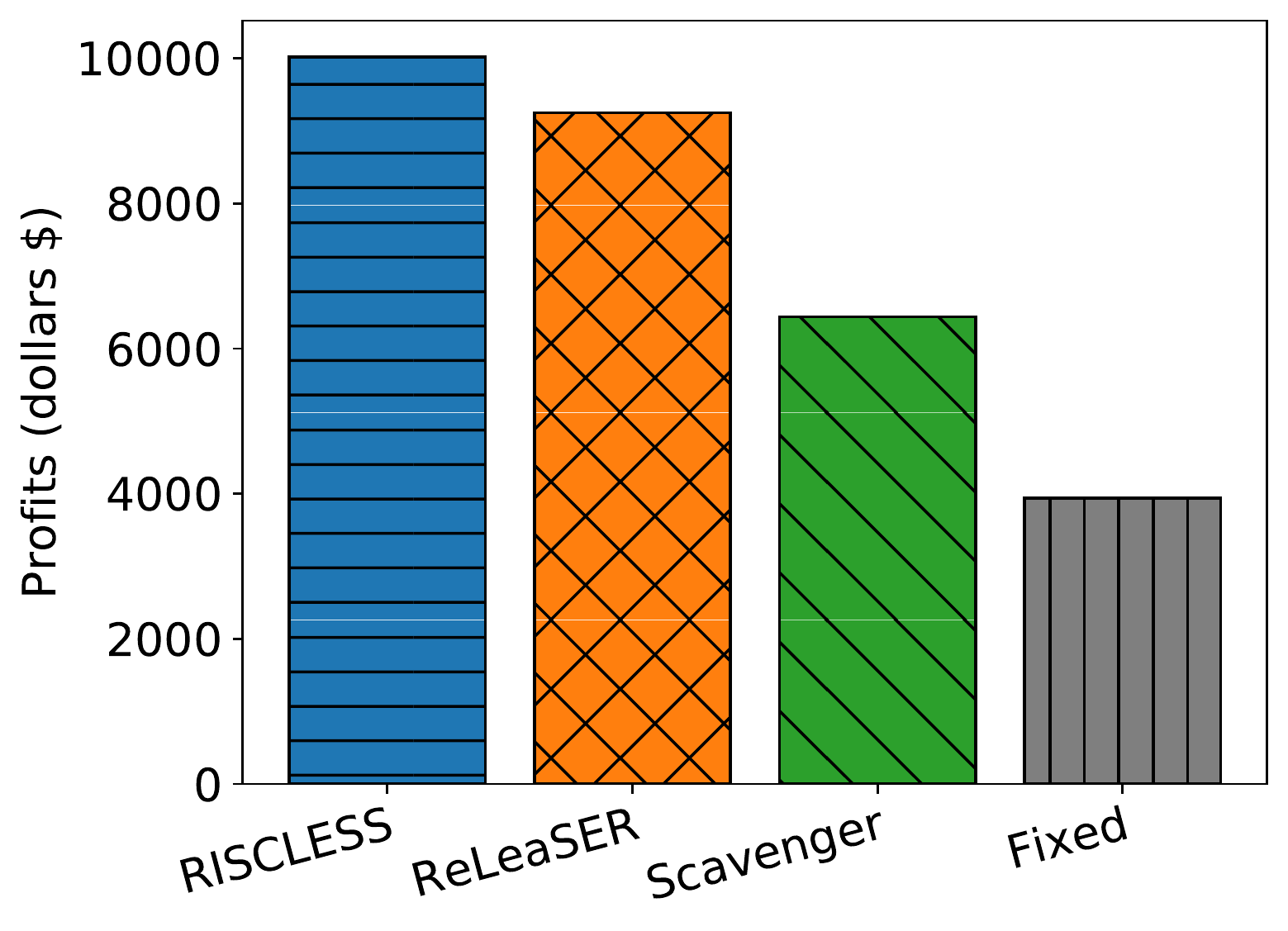}
      \caption{Private Company 2}
      \label{fig:savings_pc2}
    \end{subfigure}
    \begin{subfigure}[b]{0.32\textwidth}
      \centering
      \includegraphics[width=0.9\textwidth]{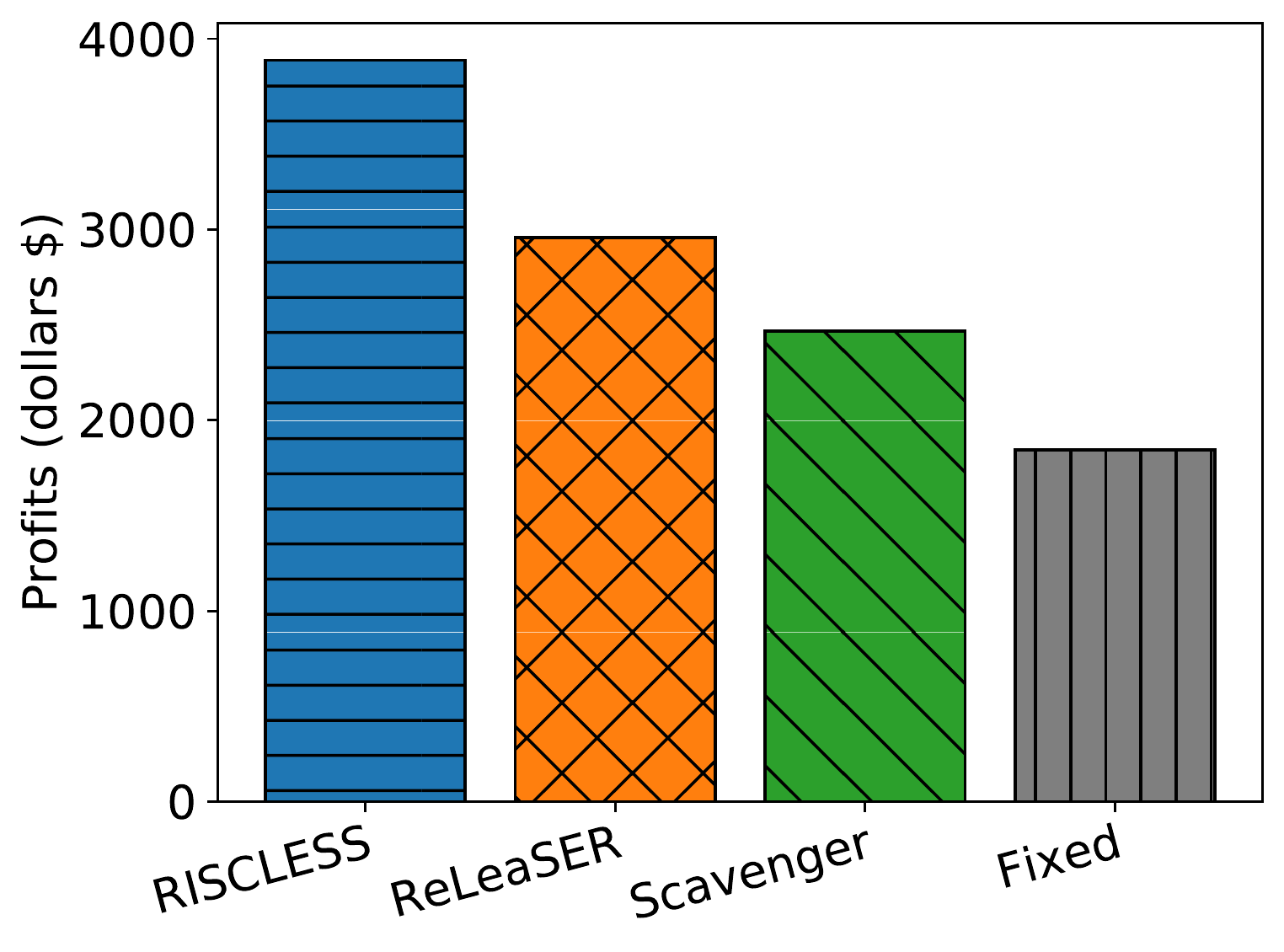}
      \caption{University}
      \label{fig:savings_university}
    \end{subfigure}
    
    \caption{Total CPs' profits over the 6-months period of traces}
    \label{fig:savings}
\end{figure*}

\textbf{1) Total profits:}
Fig.\ref{fig:savings} shows the total profits of RISCLESS and the different approaches evaluated over the 6 months traces for each datacenter.

We observe that for the three datacenters, the Fixed strategy generates the least profits for CPs. We also observe that ReLeaSER performs better than Scavenger by an average of 27.6\%. Finally, we observe that RISCLESS generates the highest profits compared to other approaches. RISCLESS improves the profits compared to ReLeaSER by 8\%, 8.3\%, and 31.5\% corresponding to PC-1, PC-2, and University respectively. These results are explained by the two following metrics namely SLA violation time and the amount of reclaimed resources.

\begin{figure*}[htbp]
    \centering
    \begin{subfigure}[b]{0.32\textwidth}
      \centering
      \includegraphics[width=0.9\textwidth]{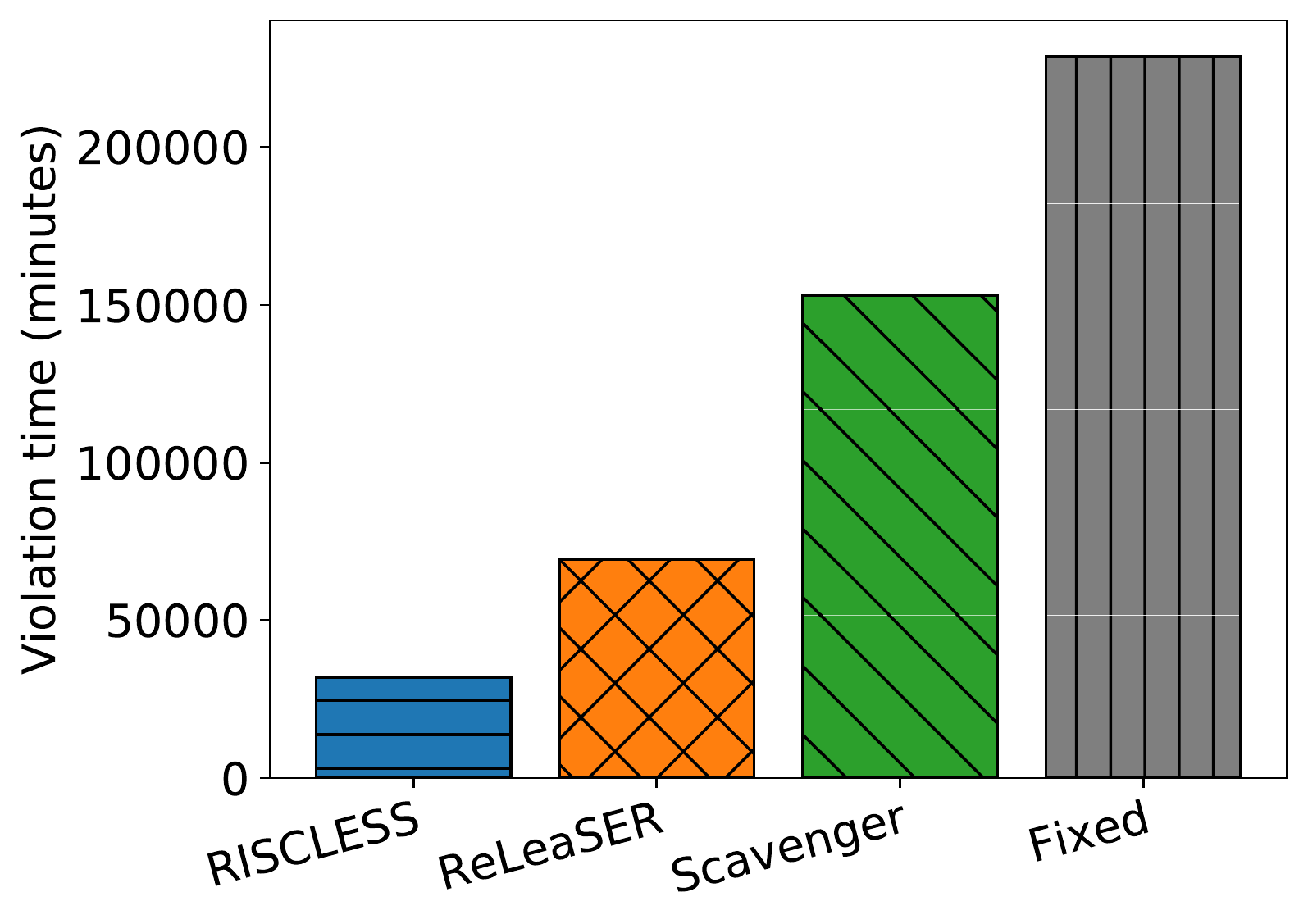}
      \caption{Private Company 1}
      \label{fig:violation_time_pc1}
    \end{subfigure}
    \begin{subfigure}[b]{0.32\textwidth}
      \centering
      \includegraphics[width=0.9\textwidth]{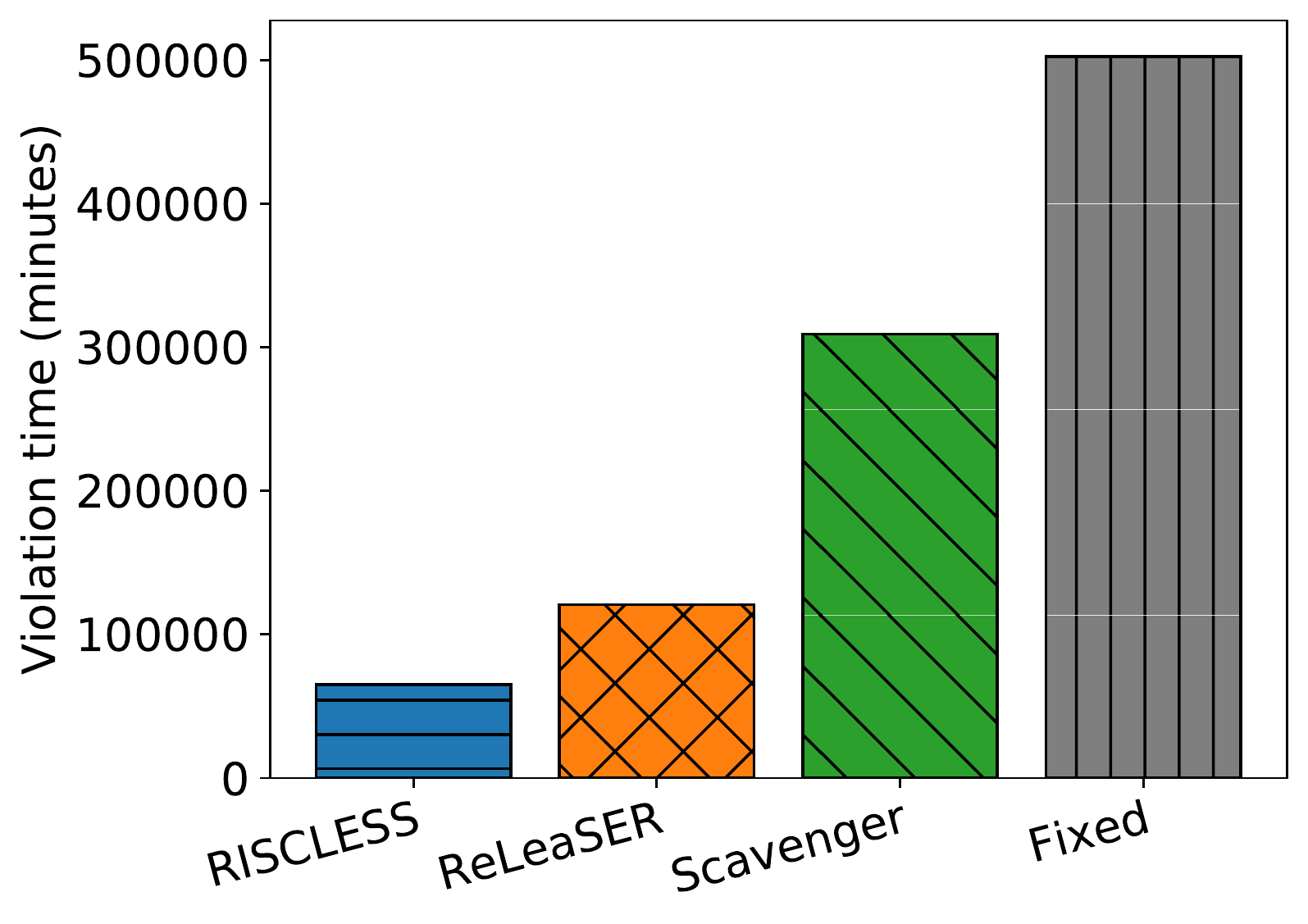}
      \caption{Private Company 2}
      \label{fig:violation_time_pc2}
    \end{subfigure}
    \begin{subfigure}[b]{0.32\textwidth}
      \centering
      \includegraphics[width=0.9\textwidth]{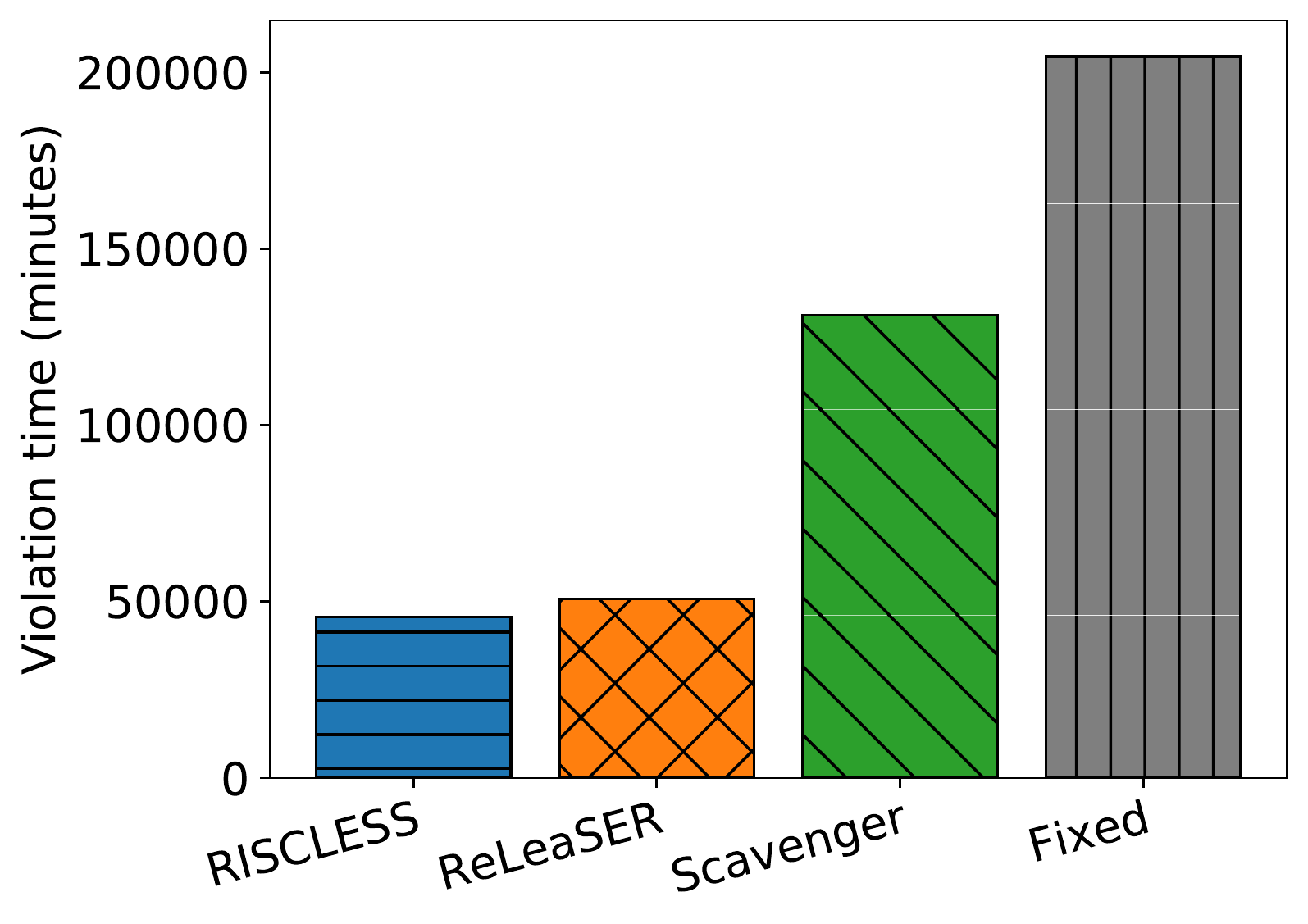}
      \caption{University}
      \label{fig:violation_time_university}
    \end{subfigure}
    
    \caption{Cumulative violation time of SLA over the 6-months period of traces}
    \label{fig:violation_time}
\end{figure*}

\textbf{2) SLA violation time:}
Fig.\ref{fig:violation_time} shows the cumulative time during which the SLA is violated. The time is summed over all days of the six months of the evaluation period for each datacenter.

We observe that RISCLESS violates SLA less than ReLeaSER, Scavenger, and the Fixed approach. RISCLESS reduces the cumulative violation time when compared to ReLeaSER by 54\%, 46.2\% and 10\% corresponding to PC-1, PC-2, and University respectively. These results show that the utilization of stable resources can decrease the SLA violation time. This partly explains the improvements in the profits seen previously.


\textbf{3) The amount of reclaimed ephemeral resources:} in this section we compute the average utilization of ephemeral resources by the different strategies. We do not take into account the ephemeral resources that were allocated but removed (\ie lost resources) because regular customers reclaimed them back. The difference between the maximum reclaimable resources and the allocated resources by a strategy in increased by the safety margin and by the removed resources. Also recall that we supposed that customers requests correspond to the maximum reclaimable resources (see Section \ref{sec:requests_assumption}).

\begin{figure*}[htbp]
    \centering
    \begin{subfigure}[b]{0.32\textwidth}
      \centering
      \includegraphics[width=0.9\textwidth]{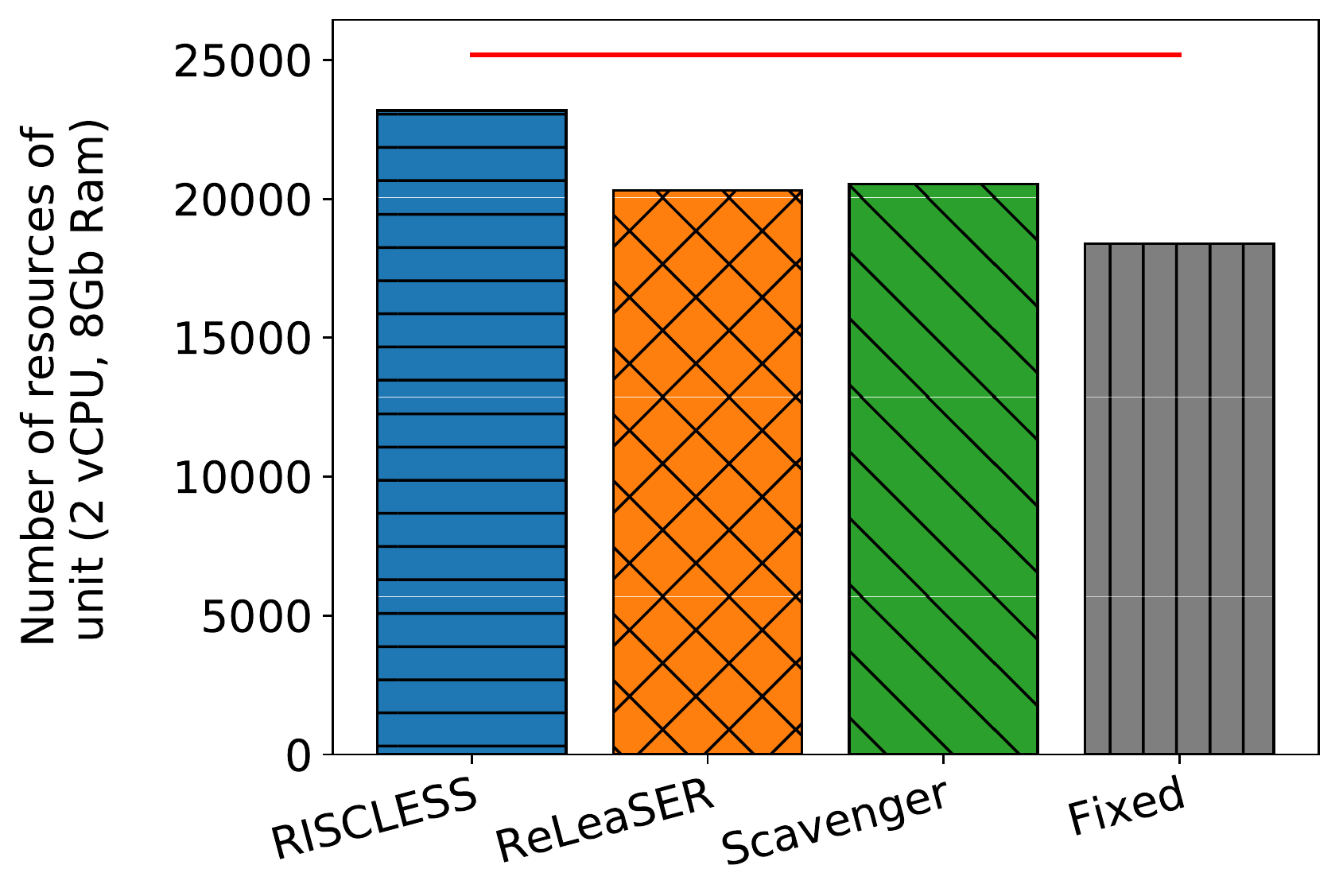}
      \caption{Private Company 1}
      \label{fig:numbre_vm_per_day_pc1}
    \end{subfigure}
    \begin{subfigure}[b]{0.32\textwidth}
      \centering
      \includegraphics[width=0.9\textwidth]{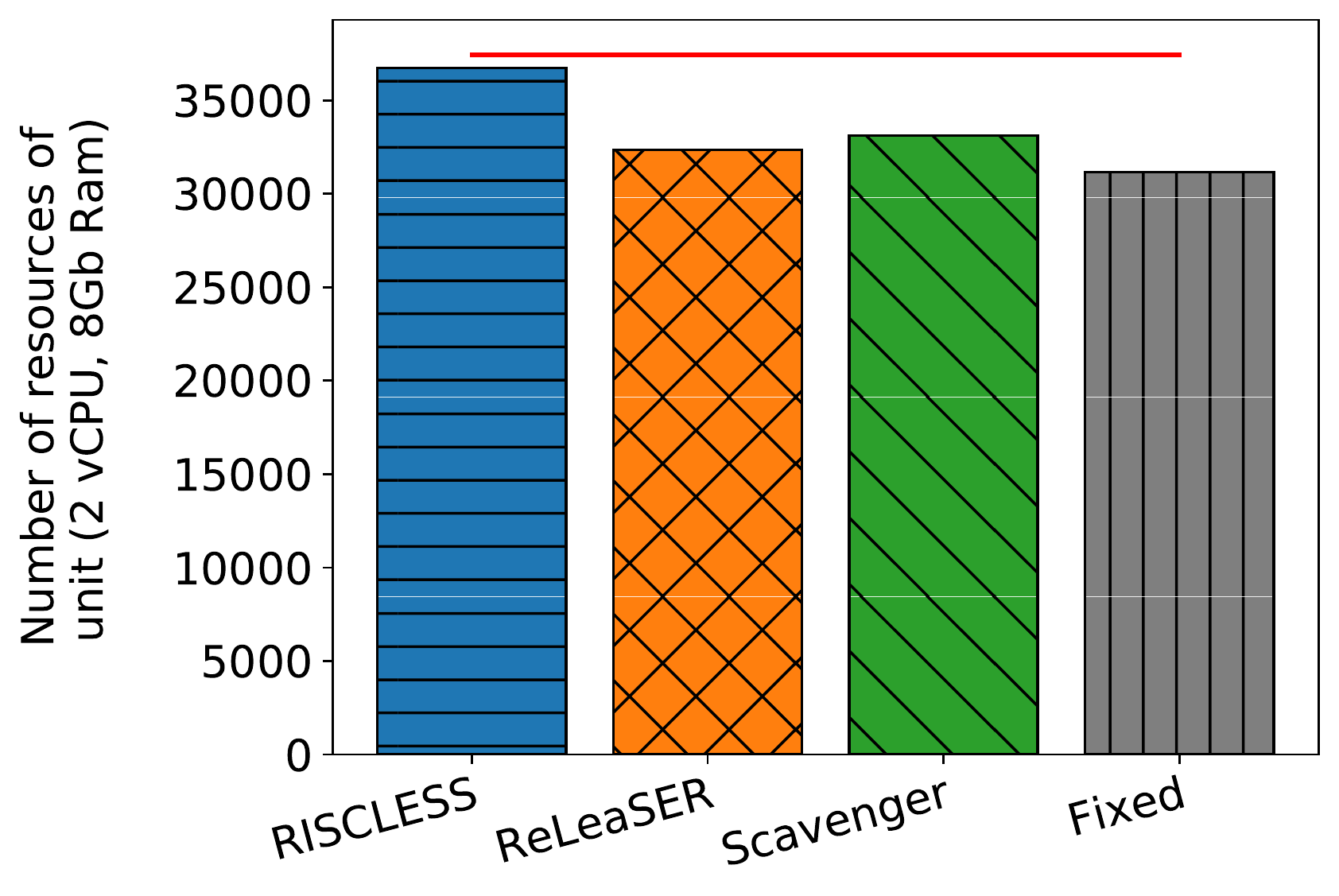}
      \caption{Private Company 2}
      \label{fig:numbre_vm_per_day_pc2}
    \end{subfigure}
    \begin{subfigure}[b]{0.32\textwidth}
      \centering
      \includegraphics[width=0.9\textwidth]{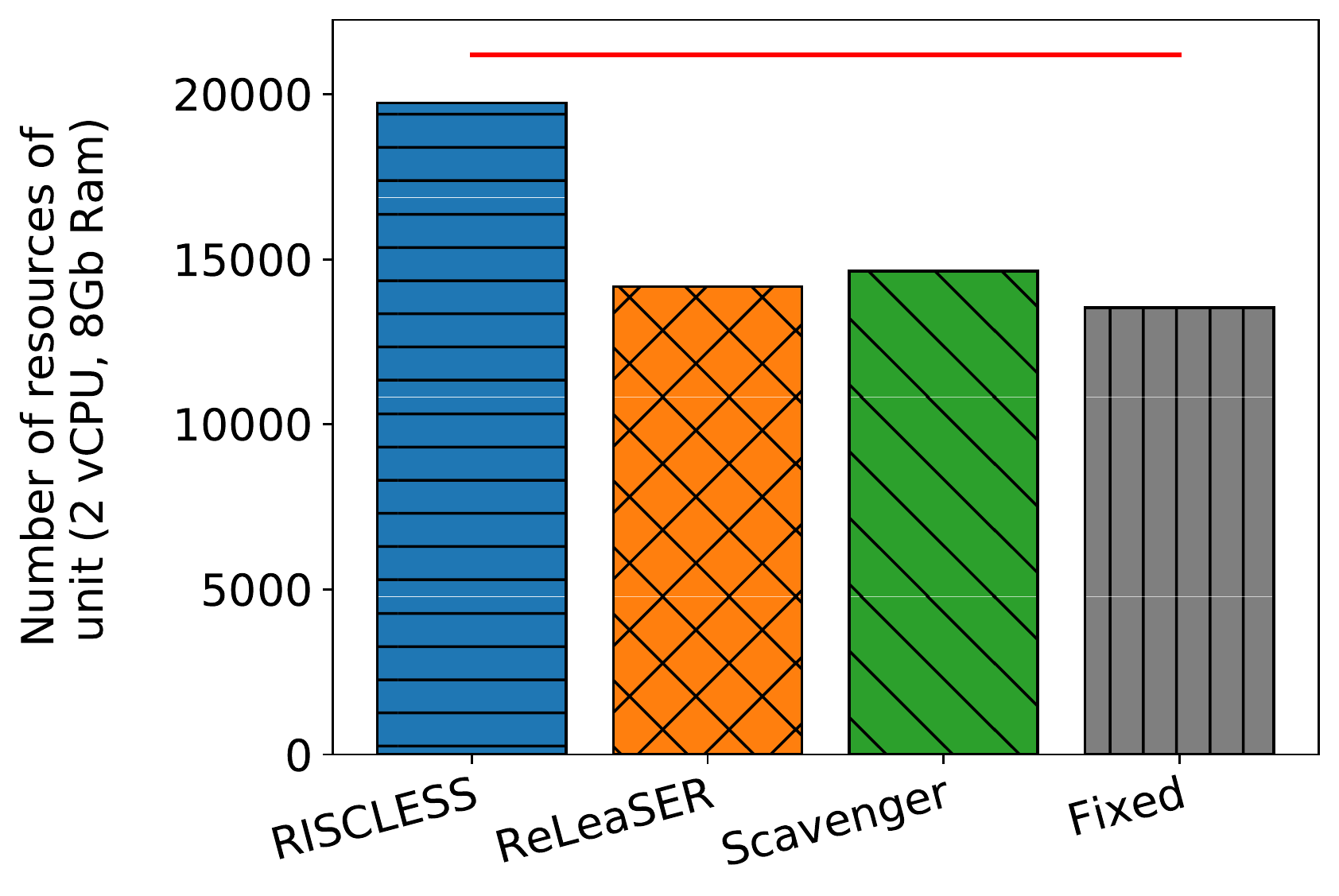}
      \caption{University}
      \label{fig:numbre_vm_per_day_university}
    \end{subfigure}
    
    \caption{Average cumulative amount of ephemeral resources used per day} 
    \label{fig:numbre_vm_per_day}
\end{figure*}

Fig.\ref{fig:numbre_vm_per_day} shows the average amount of used ephemeral resources per day (over the six months of evaluation) for each datacenter. This amount is measured as the cumulative number of allocated resource units each time step throughout the day. The red line in each figure shows the maximum reclaimable resources.

We observe that the Fixed approach utilizes the least ephemeral resources which can be explained by the safety margin used. ReLeaSER uses around 2\% fewer resources when compared to Scavenger but still manages to generate more profits. This is mainly thanks to the reduction in the SLA violation time. We also observe that RISCLESS uses more ephemeral resources per day for all datacenters. When compared to Scavenger, RISCLESS improves the utilization by 12.8\%, 10.9\%, and 34.8\% corresponding to PC-1, PC-2, and University datacenters respectively. The approaches that use a safety margin reduce the amount of usable resources (as compared to the maximum reclaimable resources) to avoid SLA violations. However, using stable resources to absorb the potential loss of volatile resources allows RISCLESS to optimize its utilization. RISCLESS uses 92\%, 98\%, and 93\% of the maximum reclaimable resources (\ie ephemeral) corresponding PC-1, PC-2, and University datacenters while having the least of SLA violations.

\vspace{.2cm}
In summary, the results of this experiment show that RISCLESS improves resource utilization and reduces SLA violations resulting in better profits for CPs. The use of on-demand stable resources allows compensating for the possible loss of the allocated ephemeral resources. This guarantees, in most cases, the SLA for customers.

\subsubsection{\textbf{RQ-2. Percentage of stable resources:}} in this section, we analyze the results from the previous experiment concerning the percentage of stable resources used. The goal is to extract some of the environment variables (\eg volatility rate) that affect the utilization of stable resources. This analysis can give CPs an idea of how many resources to reserve alongside the ephemeral ones in order to reduce SLA violations and increase their profits. Table \ref{tab:amount_stable_resources} specifies, for each datacenter, the average volatility rate per day of the ephemeral resources (computed by the Volatility Calculator), as well as the percentage of stable resources used compared to the total allocated resources.



\begin{table}
    \fontsize{10}{12}\selectfont
    \caption{The Average volatility rate and the percentage of stable resources used per day for different datacenters}
    \label{tab:amount_stable_resources}
    \begin{tabular}{cccc}
      \hline
      & PC-1 & PC-2  & University \\
      \hline
      Volatility rate & 0.83 & 0.69 & 0.75 \\
      Stable resources used (\%) & 9.21\%  & 4.63\% & 8.33\% \\
      \hline
    \end{tabular}
\end{table}


We can observe that PC-1 has the highest volatility rate and the highest percentage of stable resources. Meanwhile, PC-2 has the lowest volatility rate and stable resources.
We observe that RISCLESS uses less than 10\% of stable resources for all the tested datacenters. This translates to the use of 8 vCPU and 32 Gb of memory of stable resources on average for our datacenters (see both Table \ref{tab:datacenters_capacity} and Table \ref{tab:amount_stable_resources}). Note that this analysis can be affected by multiple variables (\ie average resource utilization, volatility rate, the number of customers' requests, etc.) and may not generalize to others. It can be observed that the more volatile the resources of a datacenter are, the more stable resources are used (however, a correlation is not conclusive from only 3 measures.). Indeed, the probability of losing ephemeral resources is expressed by their volatility rate. This rate is used by our model that decides when and how much stable resources to use in order to make up for the eventual loss of resources.


%% file: related_work.tex
\section{Related Work}
\label{section:related_work}

Previous studies about optimizing the utilization of ephemeral resources with SLA guarantees can be classified into two categories:

\begin{enumerate}[leftmargin=4ex]
    \setlength\itemsep{2ex}
    \item \textbf{Safety margin-based approaches:} safety margin was used in \cite{dartois2019cuckoo, handaoui2020salamander, zhang2016history, zhao2020rhythm, patel2020clite} with a fixed percentage of safety margin. Even though the fixed method does reduce SLA violations, it can be improved considerably alongside resource utilization since customers' workloads are not stable. Hence, a dynamic safety margin was used in Scavenger~\cite{javadi2019scavenger} and ReLeaSER~\cite{handaoui2020releaser}. It improved considerably the utilization of ephemeral resources while reducing customers' SLA violations. That being said, when resource volatility is high, the safety margin strategies may not perform well. Indeed, the higher the volatility, the larger the safety margin, the less ephemeral resources are exploited.
    
    \item \textbf{Stable and ephemeral resources:} other studies~\cite{lin2010moon, yan2016tr, yang2017pado, sharma2017portfolio, lin2020backup, ogden2019cloudcoaster} tried to improve customers SLA by utilizing stable on-demand resources on top of the ephemeral ones. The stable resources can be used for saving data in the case of data processing applications. It can also be useful for running prioritized jobs that have to be otherwise re-executed due to the lost resources. 
    Similarly, Elastigroup\footnote{\url{https://spot.io/products/elastigroup/}} is a commercialized service that offers to customers automated infrastructure scaling at a lower price by combining volatile and stable resources. However, the aforementioned solutions mainly focus on Amazon Spot Instance which is less volatile than the reclaimed resources. Furthermore, the customers using these resources receive a notification from Amazon prior to the actual interruption. This signal can be used as a convenient moment for allocating stable resources. In addition, the use cases of these solutions are generally limited to data processing applications. 
\end{enumerate}
\vspace{.2cm}
Our contribution falls into the second category. The main difference is that we operate with reclaimed resources which are highly volatile compared to Amazon Spot Instance. In our case, a notification of eviction does not exist. Thus guaranteeing SLA while maximizing CPs' profits is a challenge that we tackled using a strategy based on Reinforcement Learning. 

%% file: conclusion.tex
\section{Conclusion and Future work}
\label{section:conclusion}

In order to enhance Cloud resource utilization, CPs are leaning towards exploiting their unused resources. In this context, researchers try to find a resource allocation strategy that allows the recovery of unused resources and their exploitation without impacting customers' SLA. Although the proposed solutions do perform reasonably well, they fell short in optimizing either resource utilization or SLA guarantees. Through this paper, we proposed RISCLESS, a strategy that makes it possible to exploit ephemeral resources while reducing SLA violations. Our approach is based on RL as a decision-making model. It combines ephemeral resources with on-demand stable resources in order to offer SLA guarantees while reducing costs. The experimental evaluation results showed that RISCLESS had allowed for more thorough exploitation of ephemeral resources with a reduction in SLA violations, which significantly increased CPs' profits from the resale of resources.

As future work, we plan to include other resource metrics such as network and I/O. We also plan to improve the Volatility Calculator module to account for the amount of violated resources. Finally, we plan to incorporate Safe RL~\cite{garcia2015comprehensive} which is a technique used to improve the learning process and avoid random decisions that could have a negative impact on customers.

%% file: main.bbl
\begin{thebibliography}{10}
\providecommand{\url}[1]{#1}
\csname url@samestyle\endcsname
\providecommand{\newblock}{\relax}
\providecommand{\bibinfo}[2]{#2}
\providecommand{\BIBentrySTDinterwordspacing}{\spaceskip=0pt\relax}
\providecommand{\BIBentryALTinterwordstretchfactor}{4}
\providecommand{\BIBentryALTinterwordspacing}{\spaceskip=\fontdimen2\font plus
\BIBentryALTinterwordstretchfactor\fontdimen3\font minus
  \fontdimen4\font\relax}
\providecommand{\BIBforeignlanguage}[2]{{%
\expandafter\ifx\csname l@#1\endcsname\relax
\typeout{** WARNING: IEEEtran.bst: No hyphenation pattern has been}%
\typeout{** loaded for the language `#1'. Using the pattern for}%
\typeout{** the default language instead.}%
\else
\language=\csname l@#1\endcsname
\fi
#2}}
\providecommand{\BIBdecl}{\relax}
\BIBdecl

\bibitem{RISCLESS}
S.~Yalles, M.~Handaoui, J.-E. Dartois, O.~Barais, L.~d’Orazio, and
  J.~Boukhobza, ``Riscless: A reinforcement learning strategy to guarantee sla
  on cloud ephemeral and stable resources,'' in \emph{2022 30th Euromicro
  International Conference on Parallel, Distributed and Network-based
  Processing (PDP)}, 2022, pp. 83--87.

\bibitem{cloud_computing_nist}
N.~I. of~Standards and Technology, ``Final version of nist cloud computing
  definition published,''
  \url{https://www.nist.gov/news-events/news/2011/10/final-version-nist-cloud-computing-definition-published},
  accessed: 2020-09-16.

\bibitem{zhang2016history}
Y.~Zhang, G.~Prekas, G.~M. Fumarola, M.~Fontoura, {\'I}.~Goiri, and
  R.~Bianchini, ``History-based harvesting of spare cycles and storage in
  large-scale datacenters,'' in \emph{12th USENIX Symposium on Operating
  Systems Design and Implementation}, 2016, pp. 755--770.

\bibitem{dartois2018using}
J.-E. Dartois, A.~Knefati, J.~Boukhobza, and O.~Barais, ``Using quantile
  regression for reclaiming unused cloud resources while achieving sla,'' in
  \emph{2018 IEEE International Conference on Cloud Computing Technology and
  Science (CloudCom)}.\hskip 1em plus 0.5em minus 0.4em\relax IEEE, 2018, pp.
  89--98.

\bibitem{dartois2019cuckoo}
J.-E. Dartois, H.~B. Ribeiro, J.~Boukhobza, and O.~Barais, ``Cuckoo:
  Opportunistic mapreduce on ephemeral and heterogeneous cloud resources,'' in
  \emph{2019 IEEE 12th International Conference on Cloud Computing
  (CLOUD)}.\hskip 1em plus 0.5em minus 0.4em\relax IEEE, 2019, pp. 396--403.

\bibitem{javadi2019scavenger}
S.~A. Javadi, A.~Suresh, M.~Wajahat, and A.~Gandhi, ``Scavenger: A black-box
  batch workload resource manager for improving utilization in cloud
  environments,'' in \emph{Proceedings of the ACM Symposium on Cloud
  Computing}, 2019, pp. 272--285.

\bibitem{handaoui2020salamander}
M.~Handaoui, J.-E. Dartois, L.~Lemarchand, and J.~Boukhobza, ``Salamander: a
  holistic scheduling of mapreduce jobs on ephemeral cloud resources,'' in
  \emph{The 20th IEEE/ACM International Symposium on Cluster, Cloud and Grid
  Computing (CCGRID)}, 2020, pp. 320--329.

\bibitem{cao2014cpu}
J.~Cao, J.~Fu, M.~Li, and J.~Chen, ``Cpu load prediction for cloud environment
  based on a dynamic ensemble model,'' in \emph{Software: Practice and
  Experience}, 2014, pp. 793--804.

\bibitem{fox2009above}
A.~Fox, R.~Griffith, A.~Joseph, R.~Katz, A.~Konwinski, G.~Lee, D.~Patterson,
  A.~Rabkin, I.~Stoica \emph{et~al.}, ``Above the clouds: A berkeley view of
  cloud computing,'' in \emph{Dept. Electrical Eng. and Comput. Sciences,
  University of California, Berkeley, Rep. UCB/EECS}, 2009, p. 2009.

\bibitem{dartois2019investigating}
J.-E. Dartois, J.~Boukhobza, A.~Knefati, and O.~Barais, ``Investigating machine
  learning algorithms for modeling ssd i/o performance for container-based
  virtualization,'' in \emph{IEEE transactions on cloud computing}, 2019, pp.
  1--15.

\bibitem{zhao2020rhythm}
L.~Zhao, Y.~Yang, K.~Zhang, X.~Zhou, T.~Qiu, K.~Li, and Y.~Bao, ``Rhythm:
  component-distinguishable workload deployment in datacenters,'' in
  \emph{Proceedings of the Fifteenth European Conference on Computer Systems},
  2020, pp. 1--17.

\bibitem{patel2020clite}
T.~Patel and D.~Tiwari, ``Clite: Efficient and qos-aware co-location of
  multiple latency-critical jobs for warehouse scale computers,'' in \emph{2020
  IEEE International Symposium on High Performance Computer Architecture
  (HPCA)}.\hskip 1em plus 0.5em minus 0.4em\relax IEEE, 2020, pp. 193--206.

\bibitem{handaoui2020releaser}
M.~Handaoui, J.-E. Dartois, J.~Boukhobza, O.~Barais, and L.~d'Orazio,
  ``Releaser: A reinforcement learning strategy for optimizing utilization of
  ephemeral cloud resources,'' in \emph{2020 IEEE International Conference on
  Cloud Computing Technology and Science (CloudCom)}.\hskip 1em plus 0.5em
  minus 0.4em\relax IEEE, 2020, pp. 1--8.

\bibitem{lin2010moon}
H.~Lin, X.~Ma, J.~Archuleta, W.-c. Feng, M.~Gardner, and Z.~Zhang, ``Moon:
  Mapreduce on opportunistic environments,'' in \emph{Proceedings of the 19th
  ACM International Symposium on High Performance Distributed Computing}, 2010,
  pp. 95--106.

\bibitem{yan2016tr}
Y.~Yan, Y.~Gao, Y.~Chen, Z.~Guo, B.~Chen, and T.~Moscibroda, ``Tr-spark:
  Transient computing for big data analytics,'' in \emph{Proceedings of the
  Seventh ACM Symposium on Cloud Computing}, 2016, pp. 484--496.

\bibitem{yang2017pado}
Y.~Yang, G.-W. Kim, W.~W. Song, Y.~Lee, A.~Chung, Z.~Qian, B.~Cho, and B.-G.
  Chun, ``Pado: A data processing engine for harnessing transient resources in
  datacenters,'' in \emph{Proceedings of the Twelfth European Conference on
  Computer Systems}, 2017, pp. 575--588.

\bibitem{sharma2017portfolio}
P.~Sharma, D.~Irwin, and P.~Shenoy, ``Portfolio-driven resource management for
  transient cloud servers,'' in \emph{Proceedings of the ACM on Measurement and
  Analysis of Computing Systems}, 2017, pp. 1--23.

\bibitem{lin2020backup}
L.~Lin, L.~Pan, and S.~Liu, ``Backup or not: An online cost optimal algorithm
  for data analysis jobs using spot instances,'' in \emph{IEEE Access}, 2020,
  pp. 144\,945--144\,956.

\bibitem{ogden2019cloudcoaster}
S.~S. Ogden and T.~Guo, ``Cloudcoaster: Transient-aware bursty
  datacenterworkload scheduling,'' in \emph{arXiv preprint arXiv:1907.02162},
  2019, pp. 1--7.

\bibitem{chen2017deep}
W.~Chen, Y.~Xu, and X.~Wu, ``Deep reinforcement learning for multi-resource
  multi-machine job scheduling,'' in \emph{arXiv preprint arXiv:1711.07440},
  2017.

\bibitem{li2019deepjs}
F.~Li and B.~Hu, ``Deepjs: Job scheduling based on deep reinforcement learning
  in cloud data center,'' in \emph{Proceedings of the 2019 4th International
  Conference on Big Data and Computing}, 2019, pp. 48--53.

\bibitem{liang2019job}
S.~Liang, Z.~Yang, F.~Jin, and Y.~Chen, ``Job scheduling on data centers with
  deep reinforcement learning,'' in \emph{arXiv preprint arXiv:1909.07820},
  2019.

\bibitem{ye2018new}
Y.~Ye, X.~Ren, J.~Wang, L.~Xu, W.~Guo, W.~Huang, and W.~Tian, ``A new approach
  for resource scheduling with deep reinforcement learning,'' in \emph{arXiv
  preprint arXiv:1806.08122}, 2018.

\bibitem{dutreilh2011using}
X.~Dutreilh, S.~Kirgizov, O.~Melekhova, J.~Malenfant, N.~Rivierre, and
  I.~Truck, ``Using reinforcement learning for autonomic resource allocation in
  clouds: towards a fully automated workflow,'' in \emph{ICAS 2011, The Seventh
  International Conference on Autonomic and Autonomous Systems}, 2011, pp.
  67--74.

\bibitem{galstyan2004resource}
A.~Galstyan, K.~Czajkowski, and K.~Lerman, ``Resource allocation in the grid
  using reinforcement learning,'' in \emph{Proceedings of the Third
  International Joint Conference on Autonomous Agents and Multiagent
  Systems-Volume 3}, 2004, pp. 1314--1315.

\bibitem{liu2017hierarchical}
N.~Liu, Z.~Li, J.~Xu, Z.~Xu, S.~Lin, Q.~Qiu, J.~Tang, and Y.~Wang, ``A
  hierarchical framework of cloud resource allocation and power management
  using deep reinforcement learning,'' in \emph{2017 IEEE 37th International
  Conference on Distributed Computing Systems (ICDCS)}.\hskip 1em plus 0.5em
  minus 0.4em\relax IEEE, 2017, pp. 372--382.

\bibitem{geron2019hands}
A.~G{\'e}ron, \emph{Hands-on machine learning with Scikit-Learn, Keras, and
  TensorFlow: Concepts, tools, and techniques to build intelligent
  systems}.\hskip 1em plus 0.5em minus 0.4em\relax O'Reilly Media, 2019.

\bibitem{bellman1957markovian}
R.~Bellman, ``A markovian decision process,'' in \emph{Indiana University
  Mathematics Journal}, 1957, pp. 679--684.

\bibitem{goodfellow2016deep}
I.~Goodfellow, Y.~Bengio, A.~Courville, and Y.~Bengio, \emph{Deep
  learning}.\hskip 1em plus 0.5em minus 0.4em\relax MIT press Cambridge, 2016,
  vol.~1.

\bibitem{stigler1986history}
S.~M. Stigler, \emph{The history of statistics: The measurement of uncertainty
  before 1900}.\hskip 1em plus 0.5em minus 0.4em\relax Harvard University
  Press, 1986.

\bibitem{mnih2015human}
V.~Mnih, K.~Kavukcuoglu, D.~Silver, A.~A. Rusu, J.~Veness, M.~G. Bellemare,
  A.~Graves, M.~Riedmiller, A.~K. Fidjeland, G.~Ostrovski \emph{et~al.},
  ``Human-level control through deep reinforcement learning,'' in
  \emph{nature}, 2015, pp. 529--533.

\bibitem{chollet2015keras}
F.~Chollet \emph{et~al.}, ``Keras,'' \url{https://keras.io}, 2015.

\bibitem{abadi2016tensorflow}
M.~Abadi, A.~Agarwal, P.~Barham, E.~Brevdo, Z.~Chen, C.~Citro, G.~S. Corrado,
  A.~Davis, J.~Dean, M.~Devin \emph{et~al.}, ``Tensorflow: Large-scale machine
  learning on heterogeneous distributed systems,'' in \emph{arXiv preprint
  arXiv:1603.04467}, 2016.

\bibitem{garcia2015comprehensive}
J.~Garc{\i}a and F.~Fern{\'a}ndez, ``A comprehensive survey on safe
  reinforcement learning,'' in \emph{Journal of Machine Learning Research},
  2015, pp. 1437--1480.

\end{thebibliography}
